\documentclass{aa}
\usepackage{latexsym}
\usepackage{natbib}
\usepackage{graphicx}
\usepackage{txfonts}
\usepackage{color}
\newcommand{\sect}[1]{Sect.\,\ref{#1}}
\newcommand{\sects}[1]{Sects.\,\ref{#1}}
\newcommand{\fig}[1]{Fig.\,\ref{#1}}

\newcommand{\eqn}[1]{(\ref{#1})}
\newcommand{\app}[1]{Appendix\,\ref{#1}}
\newcommand{\dE}{{\cal{E}}}
\newcommand{\bm}[1]{\mbox{\boldmath{$#1$}\unboldmath}}
\graphicspath{{./}{figs/}}

\begin{document}

%
\title{Limitations of force-free magnetic field extrapolations: revisiting basic assumptions}

\authorrunning{H. Peter et al.}
\titlerunning{Limitations of force-free magnetic field extrapolations}

\author{H. Peter, J. Warnecke, L.P. Chitta, R.H. Cameron\inst{} 
       }

\institute{Max Planck Institute for Solar System Research,
           37077 G\"ottingen, Germany, email: peter@mps.mpg.de
}

\date{Received 24 July 2015 / Accepted 14 October 2015}

\abstract%
%
{%
Force-free extrapolations are widely used to study the magnetic field in the solar corona based on surface measurements.
}
{%
The extrapolations assume that the ratio of internal energy of the plasma to magnetic energy, the plasma-$\beta$ is negligible. 
Despite the widespread use of this assumption observations, models, and theoretical considerations show that $\beta$ is of the order of a few percent to more than 10\%, and thus not small.
We investigate what consequences this has for the reliability of extrapolation results.
}
{%
We use basic concepts starting with the force and the energy balance to infer relations between plasma-$\beta$ and free magnetic energy, to study the direction of currents in the corona with respect to the magnetic field, and to estimate the errors in the free magnetic energy by neglecting effects of the plasma ($\beta{\ll}1$).
A comparison with a 3D MHD model supports our basic considerations.
}
{%
If plasma-$\beta$ is of the order of the relative free energy (the ratio of the free magnetic energy to the total magnetic energy) then the pressure gradient can balance the Lorentz force.
This is the case in the solar corona, and therefore the currents are \emph{not} properly described.
In particular the error in terms of magnetic energy by neglecting the plasma is of the order of the free magnetic energy, so that the latter can \emph{not} be reliably determined by an extrapolation.
}
{%
While a force-free extrapolation might capture the magnetic structure and connectivity of the coronal magnetic field, the derived currents and free magnetic energy are not reliable.
Thus quantitative results of extrapolations on the location and amount of heating in the corona (through current dissipation) and on the energy storage of the magnetic field (e.g.\ for eruptive events) are limited.
}
%

\maketitle

\section{Introduction}\label{S:intro}

Measuring the magnetic field in the outer atmosphere of the Sun is very difficult and not yet achieved on a routine basis. While ideas for advanced instrumentation exist \citep[e.g.][]{Peter+al:2012}, so far we are restricted to coronagraphic observations above the limb \citep{Lin+al:2000,Lin+al:2004} or radio data \citep[e.g.][]{White:2005}, both still limited in resolution and coverage of coronal features. However, magnetic field measurement at sufficient resolution would be a key to understand the heating and dynamics of the corona, because they would provide the crucial information on currents that heat the corona through Ohmic dissipation and on the free magnetic energy to be converted during eruptive events such as flares and coronal mass ejections that govern space weather.

For the chromosphere a few such measurements exist \citep[e.g.][]{Solanki+al:2003,Lagg+al:2004}, but for the corona we mostly rely on extrapolations of the magnetic field from measurements mostly in the photosphere \citep[e.g.][]{Wiegelmann+Sakurai:2012,Wiegelmann+al:2014}.
Such extrapolations assume that the magnetic field fully dominates the plasma in the corona in terms of energy, i.e.\ that plasma-$\beta$ comparing the internal (viz.
thermal) energy of the plasma to the magnetic energy density is negligible, $\beta\ll1$, see \eqn{E:beta}. With this assumption one can build algorithms to calculate the coronal magnetic field and deduce the currents and the free magnetic energy, even though different methods do not give consistent results (see \sect{S:accuracy}).

However, observations, modelling, and theory do not really support that $\beta\ll1$. Already in his textbook \cite{Priest:1982} writes ``a magnetic field of 10\,G and a density of $10^{16}$\,m$^{-3}$ [is] characteristic of an active region'' (his Sect.\,6.4.2). Taking these numbers at face value, and employing the typical coronal temperature of 1\,MK,  one finds $\beta\approx0.35$. 
While this magnetic field estimate is realistic (cf.\ \sect{S:beta}), the density might be a factor of five to ten too high.
Still for temperatures of active region loops of about 3\,MK \citep{Reale:2014} one would then end up with a value of $\beta\gtrsim0.1$.

For the quiet Sun, \cite{Schrijver+vanBallegooijen:2005} provided evidence that in the upper atmosphere plasma-$\beta$ can be even of order unity.
While one might argue that these quiet Sun results are not directly transferable to active regions, our discussion in \sect{S:beta} will show that the quiet Sun is not just an exception. Based on published studies employing different techniques, we will conclude that $\beta$ is not necessarily small in the corona, but is of the order of several percent
to more than 10\%.

This finding, i.e.\ that plasma-$\beta$ is non-negliable, has severe consequences for force-free extrapolations of the magnetic field. This is especially true  if plasma-$\beta$  is of the same order as the free energy of the magnetic field, which is the case as we will discuss in \sect{S:free.energy}.
Actually, it is not sufficient that $\beta{\ll}1$ only on average, but this has to be fulfilled throughout the whole volume under consideration. The main consequences for force-free extrapolations of this 
are twofold. Firstly the currents do not have to be field-aligned and are not constrained (\sect{S:force}) and secondly the free magnetic energy as derived from the extrapolation is not reliable (\sect{S:energy}).
Therefore one has to be careful not to over-interpret quantitative results on heating through current dissipation or on energy storage of the magnetic field based on force-free extrapolations.

This paper is mainly relying on basic considerations concerning the force and energy balance. However, in the figures will compare these basic results to a 3D magneto-hydrodynamics (MHD) model of the corona above an evolved active region by \cite{Bingert+Peter:2011} and find that the numerical simulation is supporting our basic arguments (for details of the model we refer to the original paper). This 3D MHD model includes an energy equation accounting for radiative losses and heat conduction. Therefore the pressure derived from that model can be trusted \citep[cf.][]{Peter:2015}, which is essential when calculating plasma-$\beta$.
This trust is further supported by this type of models reproducing e.g.\ the width of coronal loops \citep{Peter+Bingert:2012}, the 3D structure of loops \citep{Bourdin+al:2013}, or the patterns of loops in an emerging active region \citep{Chen+al:2014}.

\section{Constraints on plasma-\boldmath$\beta$\unboldmath\ in the corona from observations, models, and theory}\label{S:beta}

The plasma-$\beta$ is a dimensionless number comparing the gas pressure $p$ and the  energy density of the magnetic field $B$,
\begin{equation}\label{E:beta}
\beta=\frac{p}{~B^2\,/\,(2\mu_0)~} ~,
\end{equation}
with the magnetic permeability $\mu_0$. Because the pressure essentially is the internal energy density (apart of a factor of 3/2 for a monoatomic ideal gas), $\beta$ compares the internal energy density to the magnetic energy density in a plasma.

The magnetic field in the corona is difficult to observe directly and consequently $\beta$ is not well constrained by observations. However, the technique of coronal seismology allows to get access at least to the (average) magnetic field along an oscillating coronal loop \citep{Edwin+Roberts:1983}. The applicability of this method to the real Sun and what this average value actually means have been confirmed e.g. by \cite{DeMoortel+Pascoe:2009} and \cite{Chen+Peter:2015}. To get the pressure, in studies of coronal seismology the density and temperature are derived using extreme UV spectroscopy or imaging (or in the worst case they are just assumed to be at the canonical coronal values for the density of $n{\approx}10^{9}$\,cm$^{-3}$ and the temperature of $T{\approx}1$\,MK).  Typically, in coronal seismology long loops (100\,Mm to 200\,Mm long) are studied, and  magnetic field strengths of the order of some 5\,G to 20\,G are found, which is consistent with direct coronagraphic measurements using the Zeeman effect \citep{Lin+al:2000,Lin+al:2004}.

Here we just highlight three studies, basically reflecting the findings
in coronal seismology \citep[for a review see][]{Nakariakov+Verwichte:2005}. \cite{Guo+al:2015} find in a cool (0.65\,MK) and dilute ($5{\times}10^{8}$\,cm$^{-3}$) loop system a field strength of about 8\,G, corresponding to $\beta{\approx}0.02$. In a hotter (1.05\,MK) and denser ($2{\times}10^{9}$\,cm$^{-3}$) loop
\cite{Nakariakov+Ofman:2001} find a magnetic field of about 13\,G, corresponding to $\beta{\approx}0.05$. For a larger sample of a dozen loops, \cite{White+Verwichte:2012} find a range of magnetic field strength, from about 3\,G to 20\,G. While they did not employ a density analysis, when using a typical coronal density of $10^{9}$\,cm$^{-3}$, this corresponds to values of $\beta$ in the range of 0.4 to 0.01.
So based on these observations of coronal seismology in general plasma-$\beta$ will range from a few percent to more than 10\,\%, which is not really small.

One of the major references when it comes to giving support to the assumption that $\beta$ should be small is the group of models by \cite{Gary:2001}.
Even recent reviews \citep[e.g.][]{Wiegelmann+Sakurai:2012} use this as the sole justification that $\beta$ would be small. Most quoted is Fig.\,3 of \cite{Gary:2001} that shows $\beta$ as a function of height for two models as extreme cases, one for a fieldline coming out of the middle of the umbra and one from a plage region. At a height of some 40\,Mm, i.e.\ at the apex of a typical coronal loop of 120\,Mm length, $\beta$ is 0.004 and 0.1 for the umbra and plage case, respectively. However, fieldlines from the centre of the umbra will not be associated with coronal loops, and thus the line \cite{Gary:2001}
shows for the plage region might be more typical for coronal loops. Therefore, even the study by \cite{Gary:2001}, usually used to argue for a very small $\beta$ in the corona, actually does indicate that $\beta$ is not so small.

\begin{figure}
\includegraphics[width=88mm]{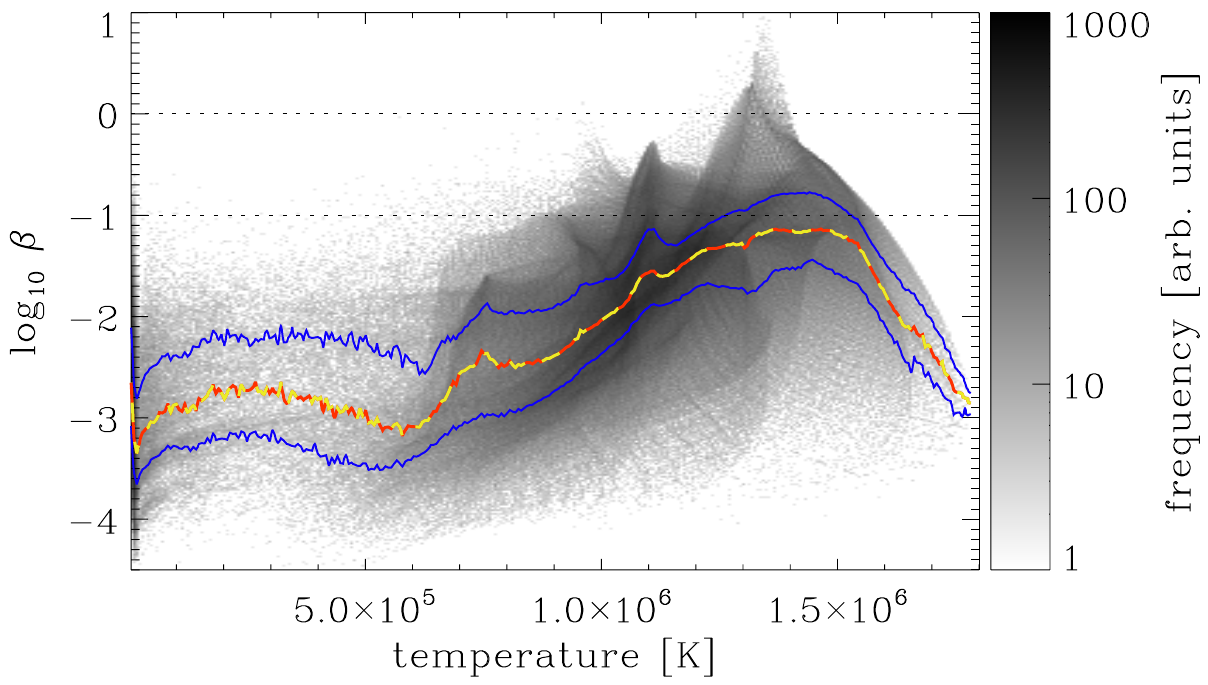}
\caption{Plasma-$\beta$ in a 3D MHD model.
The plot shows a 2D histogram of $\beta$ according to \eqn{E:beta} as a function of temperature in the the computational domain (above 5\,Mm). 
The yellow-red dashed line indicates the median variation (here similar to the mean variation). 
The solid blue lines show the 25 and 75 percentiles, i.e.\ half of the data points are in-between these lines. The color bar shows the frequency in the histogram.
Data based on the model presented in \cite{Bingert+Peter:2011}.
See \sect{S:beta}.
\label{F:beta}
}
\end{figure}

A good estimate for $\beta$ can also be inferred from 3D MHD models. If these models include a proper description of the energy equation, in particular accounting for heat conduction and radiative losses, they will give a realistic estimate for the coronal pressure and can provide a good match to the observed real Sun \citep{Peter:2015}. These models show a wide range of plasma-$\beta$ in the corona, with a median value of $\beta{\approx}0.1$ around  temperatures of about 1\,MK \citep[][Fig.\,12]{Peter+al:2006}. In places of emerging active regions where the magnetic field is comparably strong, the 3D MHD models show small values of $\beta\approx0.003$ \citep[based on data from][]{Chen+al:2015}, but in general $\beta$ is larger. In particular, \cite{Peter+al:2006} showed that in bright regions in the corona above an active region  $\beta$ is not small, but can reach values even above unity!
Small but non-negligible value of $\beta\approx0.1$ are also found in the coronal part ($T$ from 1\,MK to 1.5\,MK) of the 3D MHD model of of an evolved active region by \cite{Bingert+Peter:2011}. This and the considerable scatter of $\beta$ values in the corona is illustrated in \fig{F:beta}.

Besides the above approaches through observations and modelling, one can also estimate plasma-$\beta$ based on basic theoretical considerations by combining results from heating processes and scaling laws. In their model to heat the upper atmosphere through Alfv\'en waves, \cite{vanBallegooijen+al:2011} give a parameterisation for the volumetric heat input, $H$, depending on the magnetic field in the loop, $B$, and loop length $L$. Their Eq.\,(63) reads $H\propto B^{0.55} L^{-0.92}$.
At the same time, following \cite{Rosner+al:1978} one can find scaling relations determining the temperature $T$ and density $n$ based on $H$ and $L$. In the form of \cite{Peter+al:2012.loop} these read $n\propto H^{4/7}L^{1/7}$ and $T\propto H^{2/7}L^{4/7}$.
Combining these and using the coefficients quoted in the original papers one finds%
\begin{equation}\label{E:beta.scaling}
\beta ~=~ 0.1
       ~  \left(\frac{B}{~13\,{\rm G}~}\right)^{-1.53}
       ~  \left(\frac{L}{~100\,{\rm Mm}~}\right)^{-0.07} ~.
\end{equation}
The above mentioned values of magnetic field in coronal loops as derived from coronal seismology or coronagraphic observations using the Zeeman effect range from 5\,G to 20\,G.
Therefore with \eqn{E:beta.scaling} this consideration supports the finding that $\beta$ is \emph{not} small also from a theoretical perspective. 

The conclusion from this overview of active region observations and models is that we can expect plasma-$\beta$ to be somewhere in the range from a few percent to 10\%. 
So in contrast to commonly used statements that the corona is a low-beta plasma, we see that $\beta$ is mostly smaller than unity, but not negligibly small.
In particular the 3D MHD models show that one has to be careful with a general statement, because depending on the structure, $\beta$ might range from very small values below 0.01, to values significantly above 1, even in bright patches of the corona.
Such large values for plasma-$\beta$ have been discussed before already in the context of the quiet Sun \citep{Schrijver+vanBallegooijen:2005}, so it should not be too surprising, that also active region show non-negligible values for $\beta$.

\section{Free energy in force-free extrapolations}\label{S:free.energy}

A potential magnetic field $\bm{B}_{\rm{pot}}$ satisfies $\nabla{\cdot}\bm{B}_{\rm{pot}}{=}0$ and $\nabla{\times}\bm{B}_{\rm{pot}}{=}0$ and is a state of minimum energy under the constraint that the normal component of the field at the
photospheric boundary is prescribed.  If the magnetic field is forced away from this state, the magnetic energy available for conversion into kinetic and internal energy (i.e.\ acceleration and heating) is the free magnetic energy, $E_{\rm{free}}$. This is the difference between the actual magnetic energy, $E_{\rm{mag}}$ and the magnetic energy of the potential field (with the same boundary conditions for the normal component of $\bm{B}$), $E_{\rm{pot}}$. 
Here the energies are integrated over a volume, e.g. a full active region under investigation.

Because the spatial distribution of  the magnetic field cannot be measured directly in the corona (with some exceptions), mainly (non-linear) force-free extrapolations are employed to estimate the magnetic energy, $E_{\rm{mag}}$ (see \sect{S:force}).
In the following we give a by far non-exhaustive list of examples for such studies that provide information on the free energy. For these we give values for the relative free energy $\dE_{\rm{free}}$ in percent, where
\begin{equation}\label{E:delta.free}
\dE_{\rm{free}} ~=~ \frac{~E_{\rm{mag}}-E_{\rm{pot}}~}{E_{\rm{mag}}} ~.
\end{equation}
Here the index ``free'' refers to \emph{free energy} and not to force-free extrapolation.
This quantity is positive because for a given normal component of the magnetic field at the boundary the potential field is the solution with the minimum magnetic energy.

Force-free extrapolations evaluate this free energy in a given volume, and of course depending on the volume one would find different values. Representing the huge variety of force-free models \citep[see e.g.\ review by][]{Wiegelmann+Sakurai:2012} we will quote the results from extrapolations of three different structures: a normal (non-flaring) active region, a flaring active region, and a small highly twisted part of an active region. %
The maximum magnetic energy in a volume for a specified magnetic field at its lower boundary is given by the energy of an open field configuration. Its maximum free magnetic energy $\dE_{\rm{free}}$ is about 100\%, and this maximum also applies to force-free extrapolations \citep[e.g.][]{Amari+al:2000}.
While exceptions of large free magnetic energies exist \citep[e.g. up to 60\% in a flaring active region;][]{Thalmann+Wiegelmann:2008}, typically force-free extrapolations all show relative free energies of some 10\%.

As a sample for a normal active region we quote \cite{Thalmann+al:2012} who find a free magnetic energy $\dE_{\rm{free}}$ of about 7\% to 14\%. The domain they investigated covers about 180$\times$98\,Mm$^2$ in the horizontal directions, basically encompassing the whole active region. One might expect a higher level of free energy for a flaring active region just before a flare. While \cite{Sun+al:2012} find a free energy of almost 50\% two days before a X-class flare, just a few hours before the flare the free energy was at only 20\% (the authors did not specify the volume but it probably encompasses the whole active region).
Thus even in a flaring active region the level of free energy is modest. This is also true for a twisted subvolume in an active region. \cite{Thalmann+al:2014} investigated the region of strong braiding identified in the images of the High-resolution Coronal Imager (Hi-C) rocket flight \citep{Cirtain+al:2013}. However, even in a small volume of 13$\times$7$\times$4\,Mm$^3$ they found the free energy $\dE_{\rm{free}}$ of the order of only 5\%.
These low values for the free energy are not simply an artefact of one particular method. Comparing  major codes for force-free extrapolations \cite{DeRosa+al:2009} found that they all give small values in the range of 5\% to 25\% (full active region, 90$\times$90\,Mm$^2$ in the horizontal directions).

These values for the free magnetic energy in force-free extrapolations are also found in 3D MHD simulations. In the model of an evolved active region by \cite{Bingert+Peter:2011} that we use as an example throughout this paper we find a free magnetic energy of about 10\% when integrating over the whole computational domain covering the active region (here the free energy is dominated by the lower part of the computational box).

The conclusion from this overview of extrapolation  models is that the (relative) free magnetic energy in the corona ranges mostly from 5\% to 10\% in an active region. This might be slightly higher in flaring regions, but probably not (much) higher than 20\%.
These values should be considered as upper limits, because they depend on the volume selected for integrating the energies. When extending the volume to include, e.g. neighbouring non-active parts, where the free magnetic energy will be lower, the (relative) free energy of the whole volume will go down.

\section{Local force balance and currents}\label{S:force}

We will first check what role the plasma-$\beta$ plays in the force balance and what implications this has for the currents derived from a force-free extrapolation. This is a local treatment in the sense that we investigate the forces at a given spot somewhere in the corona,
e.g. at the location of a bright coronal loop. 

The ideal momentum balance in a static case reads
\def\rrr{\rule[-2.5ex]{0pt}{2ex}}
\begin{equation}\label{E:force}
0 ~~=~~ \underbrace{\rrr-\nabla{p}}_{\rm(a)}
  ~~+~~ \underbrace{\rrr\rho\bm{g}}_{\rm(b)}
  ~~-~~ \underbrace{\rrr\nabla\frac{B^2}{~2\mu_0~}}_{\rm(c)}
  ~~+~~ \underbrace{\rrr\frac{1}{\mu_0}\,(\bm{B}\cdot\nabla)\bm{B}}_{\rm(d)} ~.
\end{equation}
Here $p$ is the pressure, $\rho$ the density, $\bm{g}$ the gravitational acceleration, $\bm{B}$ the magnetic field, and $B=|\bm{B}|$.
We separated the Lorentz force $\bm{f}_{\rm{L}}{=}\,\bm{j}{\times}\bm{B}$ into the magnetic pressure gradient and the magnetic tension, i.e. terms (c) and (d).

The plasma-$\beta$ introduced in \eqn{E:beta} compares the gas and the magnetic pressure, i.e.\ basically terms (a) and (c,d) in \eqn{E:force}:  In general in a low-$\beta$ environment the magnetic force will dominate the pressure gradient.%
\footnote{However, if $B{=}B_0{+}B_1$ and $p{=}p_0{+}p_1$, $B_0$ and $p_0$ are constant in space, and $B_1{\ll}B_0$ and $p_1{\ll}p_0,$ even for $\beta{\ll}1$ it might well be that the pressure gradient equals the Lorentz force. This is because in this case the forces, i.e. the gradients, depend only on $B_1$ and $p_1$, while $\beta$ mainly compares $B_0$ and $p_0$.}  
In the stationary case in most situations a hydrostatic equilibrium will govern the variation along magnetic field lines (at least for structures with constant cross section), so the terms (a) and (b) in \eqn{E:force} will be of the same order, implying that the gravitational term (b) would also be small if $\beta$ is small.

Therefore in a low-$\beta$ regime the gradient of the magnetic pressure (c) has to be balanced by the magnetic tension force (d) alone. This is also true when flows are present as long as these are subsonic, because then the kinetic energy is still even smaller than the internal energy.
Consequently we have
\begin{equation}\label{E:ff.cond}
\nabla\frac{B^2}{~2\mu_0~} \approx \frac{1}{\mu_0}\,(\bm{B}\cdot\nabla)\bm{B}
\quad\Leftrightarrow\quad
\bm{j}\times\bm{B}\approx0 ~.
\end{equation}
This condition applies only approximately, i.e. to the order of $\beta$, where we still have to define what this means (see \sect{S:small.j}).
Because this implies that all forces are small, solutions to this state are called force-free.

To satisfy the condition \eqn{E:ff.cond} two (not mutually exclusive) options exist, 
\begin{equation}\label{E:ff.options}
\begin{array}{l}
\mbox{(a):}~~~\bm{j}~\mbox{is roughly parallel to}~\bm{B} ~,
\\[0.7ex]
\mbox{(b):}~~~\bm{j}~\mbox{is small, but not necessarily aligned with}~\bm{B}~.
\end{array}
\end{equation}

Seeking force-free solutions of \eqn{E:ff.cond}, research focuses on option (\ref{E:ff.options}a), 
as e.g.\ outlined in the introduction of the review by \cite{Wiegelmann+Sakurai:2012}.
For such force-free extrapolations the possibility (\ref{E:ff.options}b) of small currents not aligned with the magnetic field is largely ignored. 

However, because the condition \eqn{E:ff.cond}, $\bm{j}{\times}\bm{B}{\approx}0$, has to be fulfilled only approximately to the extent that the gas can balance the magnetic forces to the order of $\beta$, one can have either large currents at small angles (between $\bm{j}$ and $\bm{B}$), or one could have small currents at arbitrary angles. The latter solution is not the same as a potential field where $\bm{j}{=}0$ and the Lorentz force vanishes. Instead we have seen in \sect{S:beta} that plasma-$\beta$ is of the order of up to 10\%, and can in some places of the bright corona even reach values above 1! \citep[See \sect{S:beta} and][]{Peter+al:2006}. Thus there can be non-negligable Lorentz forces in the corona.

The main implication following from the two options in \eqn{E:ff.options} is that they can lead to significantly different results for the currents in the system, and thus for the expected heating rates derived from an extrapolation. Likewise, if the currents are quite different between the two options, also the free magnetic energy for these two options will be quite different.
Therefore selecting option (\ref{E:ff.options}a), $\bm{j}\,||\,\bm{B}$, makes a choice that might provide answers in terms of currents and free magnetic energy that are far from being a unique answer.

\subsection{What does ``small currents'' mean?}\label{S:small.j}

We now investigate to what extent the plasma can balance the Lorentz forces due to small currents that are \emph{not} parallel to the magnetic field. This is the same as to ask when the option (\ref{E:ff.options}b) will apply, i.e. when the currents are small.

For this we decompose the magnetic field into a part that is a potential field $\bm{B}_{\rm{pot}}$ (for the given boundary conditions) and a second part that we call the free magnetic field, $\bm{B}_{\rm{free}}$,
\begin{equation}\label{E:free.B}
\bm{B} = \bm{B}_{\rm{pot}} + \bm{B}_{\rm{free}} ~.
\end{equation}
Similar to \eqn{E:delta.free}, here the index ``free'' refers to the magnetic field corresponding to the \emph{free} magnetic energy, and not to the force-free extrapolation.

The free magnetic field $\bm{B}_{\rm{free}}$ corresponds to the current system that is not necessarily parallel to the magnetic field.
By definition we have $\nabla{\times}\bm{B}_{\rm{pot}}{=}0$, and the currents are given by $\bm{j}=(\nabla{\times}\bm{B}_{\rm{free}})/\mu_0$.
Since here we consider case (\ref{E:ff.options}b), we assume that these currents
\bm{j} are not (predominantly) aligned with the magnetic field.

Considering only the pressure gradient and the Lorentz force (terms a,c,d) we can write the force balance \eqn{E:force} as
\begin{equation}\label{E.force.free}
\nabla p = \frac{1}{\mu_0}~(\nabla\times\bm{B}_{\rm{free}})\times\bm{B} ~.
\end{equation}
Dividing this by the magnetic energy density $B^2/(2\mu_0)$ and because here we investigate the small currents not aligned with the magnetic field, we find the order of magnitude estimation
\begin{equation}\label{E:Bfree.beta}
\frac{~B_{\rm{free}}~}{B} \approx \frac{1}{2}~\beta ~.
\end{equation}

As long as $B_{\rm{free}}$ is smaller than $B_{\rm{pot}}$, the free magnetic energy is ${\approx}2\,B_{\rm{free}}B_{\rm{pot}}/(2\mu_0)$.
For this local analysis of the force balance we will consider the relative free energy $\dE_{\rm{free}}$ as defined in \eqn{E:delta.free} locally, i.e.\ we evaluate it in a small volume around the location where $B_{\rm{free}}$ and $B_{\rm{pot}}$ are taken. For an order-of-magnitude estimation we only consider the absolute value, because locally $\dE_{\rm{free}}$ can be negative (see also \sect{S:free.mhd}).
With this it follows from \eqn{E:delta.free} that $B_{\rm{free}}/B\approx|\dE_{\rm{free}}|/2$. Thus from \eqn{E:Bfree.beta} we find the relation between the relative free energy and plasma-$\beta$ in  order to satisfy option  (\ref{E:ff.options}b),
\begin{equation}\label{E:Efree.beta}
|\dE_{\rm{free}}| \approx \beta ~.
\end{equation}
This relation, based on a local force balance, is an order-of-magnitude estimation only. The comparison with a 3D MHD model in \sect{S:free.mhd} shows that there $|\dE_{\rm{free}}|$ is larger than $\beta$, typically by a factor of about five.

From the relation in \eqn{E:Efree.beta} it follows that as long as the relative free energy $\dE_{\rm{free}}$ \eqn{E:delta.free} is of the order of plasma-$\beta$ (or smaller), the currents not parallel to the magnetic field are small enough, so that the plasma can balance the resulting Lorentz force.
In this case, even the parallel currents are not constrained, and a force-free extrapolation would not capture the component of the (small) currents that is parallel to the magnetic field (see \app{S:para.currents}).

Only if the relative free energy would be significantly larger than plasma-$\beta$, 
the corresponding currents must be dominated by those corresponding to the non-linear force-free modelling.
Then option (\ref{E:ff.options}a) would apply and the currents would have to be (roughly) parallel to the magnetic field.

However, from the discussion in \sects{S:beta} and \ref{S:free.energy} and the MHD model in \fig{F:free.energy} it is clear, that the relative free magnetic energy is of the order of plasma-$\beta$, so that the condition \eqn{E:Efree.beta} is satisfied. 
From this we conclude, that (in most cases) the free energy is consistent with currents \emph{not} 
parallel to the magnetic field. 
Therefore the currents are not necessarily parallel to $\bm{B}$, and consequently the currents derived from (most) force-free extrapolations are only reliable in stating that the field is close to potential. Therefore also estimations of e.g.\ the localisation of plasma heating 
based on the currents from force-free extrapolations are not reliable.

\subsection{Free energy and plasma-beta in a 3D MHD simulation}\label{S:free.mhd}

\begin{figure}
\includegraphics[width=88mm]{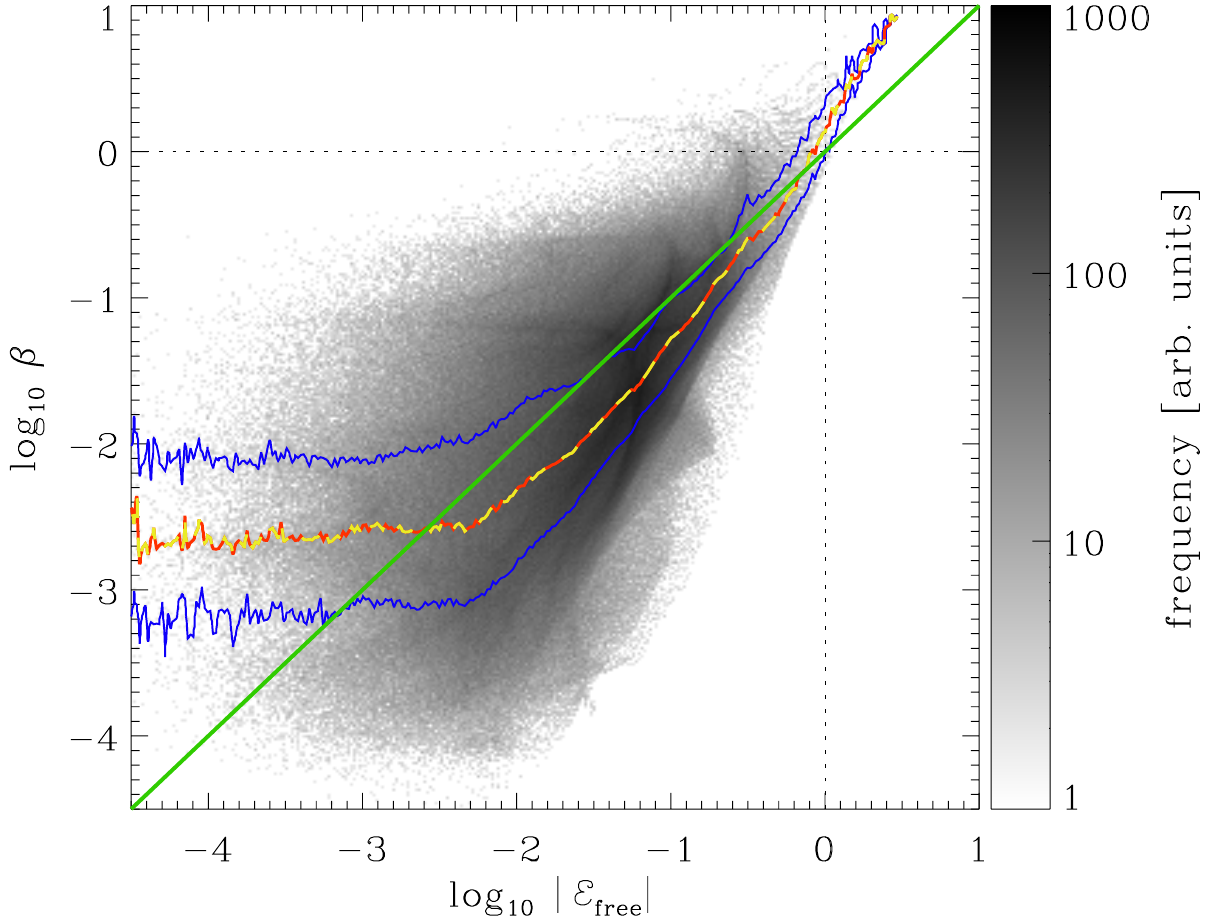}
\caption{Relation between plasma-$\beta$ and relative free magnetic energy $\dE_{\rm{free}}$ in a 3D MHD model.
For the 2D histogram plasma-$\beta$ \eqn{E:beta} is calculated at each grid cell of the computational domain (above 5\,Mm) and the free energy $\dE_{\rm{free}}$ is evaluated according to \eqn{E:delta.free} in the volume of the same respective cell. 
Here we plot the absolute values, $|\dE_{\rm{free}}|$.
The potential field required for this is calculated using the vertical component of the magnetic field in the 3D model at a height of 5\,Mm which is (roughly) the base of the corona in that model.
The yellow-red dashed line indicates the median variation (here similar to
the mean variation). 
The solid blue lines show the 25 and 75 percentiles, i.e.\ half of the data points
are in-between these lines.
For comparison the straight green line shows a linear relation $\beta\,{=}\,\dE_{\rm{free}}$. 
The color bar shows the frequency in the 2D histogram.
Data based on the model presented in \cite{Bingert+Peter:2011}.
See \sect{S:free.mhd}.
\label{F:free.energy}
}
\end{figure}

To illustrate the relation between the (local) relative free energy $\dE_{\rm{free}}$ and plasma-$\beta$ as derived in \eqn{E:Efree.beta} we show a 2D histogram of this based on the 3D MHD model of \cite{Bingert+Peter:2011} in \fig{F:free.energy}.
Just as with \eqn{E:Efree.beta}, here $\dE_{\rm{free}}$ is evaluated in one grid cell of the simulation when compared to $\beta$ at that same grid cell. Like for the order-of-magnitude estimation above, we plot the absolute values of $\dE_{\rm{free}}$.
The respective histograms for the positive and negative values alone look basically identical to the one for absolute values displayed in \fig{F:free.energy}.

The histogram in \fig{F:free.energy} shows that typically $|\dE_{\rm{free}}|$ is several (typically about five) times larger than $\beta$ (most entries in the histogram are to the right of the green line indicating $|\dE_{\rm{free}}|{=}\beta$).
Furthermore, the volume with positive values of $\dE_{\rm{free}}$ is about a factor of 1.5 larger than the one with negative values.
Consequently, when considering the whole volume $\dE_{\rm{free}}$ is positive, as it should be.
As mentioned already in \sect{S:free.energy}, for this model we find $\dE_{\rm{free}}{\approx}0.1$ when considering the whole domain.

In general the 3D MHD model supports the order-of-magnitude estimation in \eqn{E:Efree.beta}, even though it clearly shows that mostly $|\dE_{\rm{free}}|$ is larger than $\beta$.

\subsection{Currents in a 3D MHD simulation}\label{S:j.mhd}

The relation of the currents to the magnetic field, i.e. the two options  (\ref{E:ff.options}a) and (\ref{E:ff.options}b) can be directly investigated in a 3D MHD model. In such a model also the plasma is considered, i.e.\ the model fully accounts for the fact that $\beta$ is not small in general. In the 3D MHD model by \cite{Bingert+Peter:2011} we employ here, the free magnetic energy integrated over the whole volume is about 10\%. This is comparable to the values found for typical solar active regions based on force-free extrapolations (cf.\ \sect{S:free.energy}) and thus this active region models should be a good representation of the real Sun in terms of the available free magnetic energy and currents.

\begin{figure}
\includegraphics[width=88mm]{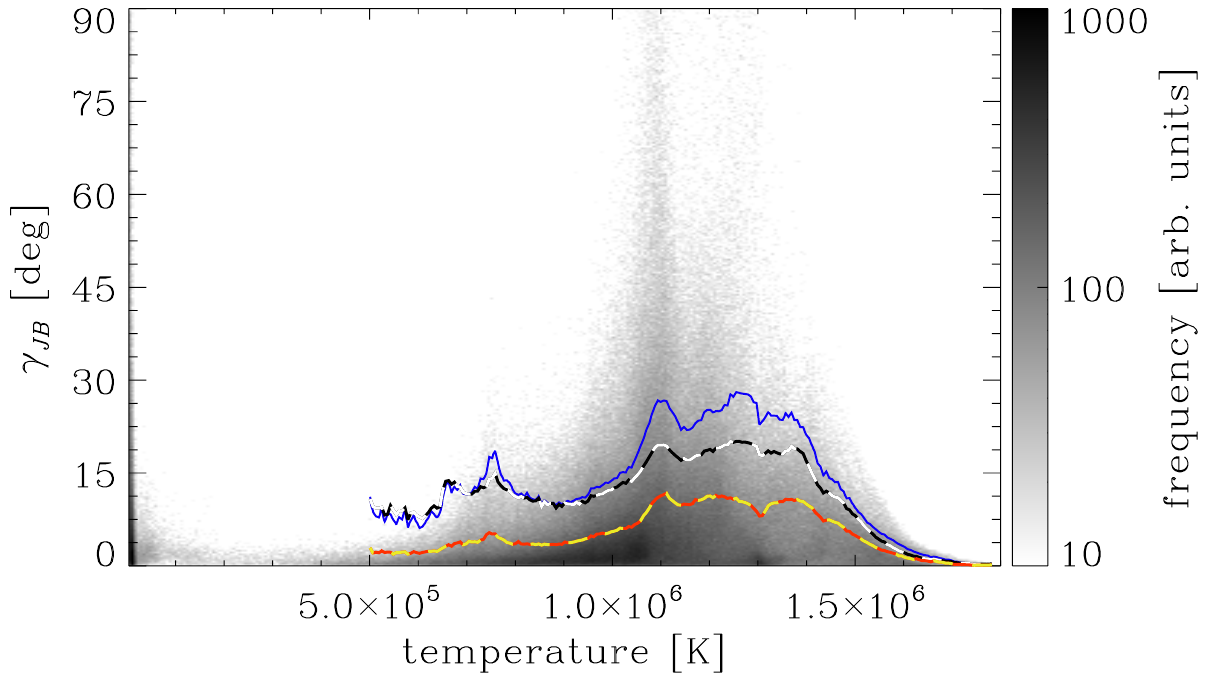}
\caption{Angle $\gamma_{JB}$ between current and magnetic field in a 3D MHD model. The plot shows a 2D histogram of the angle as a function of temperature in the computational domain (above 5\,Mm).
The yellow-red dashed line indicates the median variation, the white-black dashed line the mean variation. 
The solid blue line shows the 75 percentile, i.e.\ still 25\% of all data points have angles above the blue line. The color bar shows the frequency in the histogram.
Data based on the model presented in \cite{Bingert+Peter:2011}.
See \sect{S:j.mhd}.
\label{F:angle}
}
\end{figure}

In \fig{F:angle} we show the distribution of the angle $\gamma_{\!JB}$ between the current $\bm{j}$ and the magnetic field $\bm{B}$ in that 3D MHD model. While in general the angles are small, the mean angle in the coronal part of the computational domain around $T{\approx}1$\,MK is about 15$^\circ$, and 25\% of the coronal volume (from $T{\approx}1$\,MK to 1.5\,MK) has angles above 30$^\circ$. 
In the corona we even find angles of almost 90$^\circ$, i.e.\ currents that are almost perpendicular to the magnetic field.

This comparison to a 3D MHD models confirms that the currents in the corona are not necessarily parallel to the magnetic field as it would be required for option (\ref{E:ff.options}a) when assuming that $\beta$ is very small. Instead, when accounting for the effects of the plasma, $\beta$ is not necessarily small (cf.\ \sect{S:beta}, \fig{F:beta}) and the currents will not be parallel to the magnetic field, inline with option
 (\ref{E:ff.options}b).

\section{Global energy considerations and accuracy}\label{S:energy}

In contrast to the local treatment discussed in \sect{S:force}, the following considerations apply to the energy contained in a volume. In this sense this is non-local, or global.
As outlined in \sect{S:free.energy} the volume considered by force-free extrapolations typically ranges from a several Mm$^3$ to the better part of an active region.

Based on the virial theorem \citep[e.g.][Sect 2.8.4]{Priest:1982} in a stationary state the energy balance following from the momentum balance requires that the total energy within a volume, $E_{\rm{tot}}$, is solely determined by an integral of properties at its surface, $S$,
\begin{equation}\label{E:Virial}
0 = E_{\rm{tot}}  + S ~.
\end{equation}
As before an uppercase $E$ refers to the volume integral of the respective energy density. 
The exact form of $S$ is not of interest here, but it can be found e.g. in Eq.\,2.66 in \cite{Priest:1982}.

For the following we will consider only the magnetic and the internal (viz. thermal) energy,
\begin{equation}\label{E:Etot}
E_{\rm{tot}}  = E_{\rm{mag}} + E_{\rm{int}} ~.
\end{equation}
Other terms such as the kinetic or gravitational energy, $E_{\rm{kin}}$ or $E_{\rm{grav}}$, will only strengthen the arguments given below, because they will increase the non-magnetic contributions to the total energy.
In general, we will have $E_{\rm{kin}}\lesssim E_{\rm{int}}$ because the flows are mostly subsonic. In most situations $E_{\rm{grav}}$ will be of the order of $E_{\rm{int}}$ reflecting hydrostatic equilibrium (along fieldlines).

We rewrite \eqn{E:Virial} to give
\begin{equation}\label{E:E.balance}
0 = E_{\rm{mag}} + E_{\rm{int}} + S ~
\end{equation}
and introduce 
\begin{equation}\label{E:epsilon}
\delta = \frac{E_{\rm{int}}}{~E_{\rm{mag}}~} ~.
\end{equation}
Because the internal energy basically is the pressure, this quantity will be of the order of the (average) plasma-$\beta$ \eqn{E:beta},
\begin{equation}\label{E:epsilon.beta}
\delta \approx \frac{3}{2}~\beta ~.
\end{equation}
With \eqn{E:epsilon} we can rewrite \eqn{E:E.balance} to give
\begin{equation}\label{E:balance.epsi}
0 = E_{\rm{mag}} (1+\delta) + S ~.
\end{equation}

For a (non-linear) force-free extrapolation, one assumes that the magnetic field has to balance itself, so other forces (and energies) are negligible and thus $\delta=0$.
Consequently, the minimum magnetic energy of the force-free extrapolation, $E_{\rm{ff}}$, is given by the surface term,
\begin{equation}\label{E:energy.ff}
E_{\rm{ff}} = -S ~.
\end{equation}
Comparing with \eqn{E:balance.epsi} and using \eqn{E:epsilon.beta} we find the relative error of the force-free extrapolation (in terms of energy) with respect to the true magnetic field to be

\begin{equation}\label{E:err.ff}
\dE_{\rm{err}} 
~~=~~
\frac{E_{\rm{mag}}-E_{\rm{ff}}~}{E_{\rm{mag}}}
~~\approx~~ \frac{3}{2}~\beta ~.
\end{equation}

This implies that we can determine the magnetic energy through a force-free extrapolation only to the order of plasma-$\beta$. Because force-free extrapolations neglect the existence of the plasma, it is clear that this should be a limiting factor for the accuracy of the magnetic energy.
The above estimate is a conservative estimate, because also the gravitational and kinetic energy will contribute to this uncertainty, so that the true error could easily be also a factor of 2 to 3 larger than estimated in \eqn{E:err.ff}.

If $\beta\to0$, which is equivalent to fully ignore the effects of the plasma in the corona, the magnetic field will be in a truly force-free state. Then the accuracy is limited only by the accuracy of the algorithm used for the force-free extrapolation (\sect{S:accuracy}). 
However, on the real Sun plasma-$\beta$ is not arbitrarily small. Instead, observations, theory, and numerical simulations show values for $\beta$ in the corona ranging from a few to 10\% (see \sect{S:beta}).
Therefore we can expect the magnetic field deduced from extrapolations to be accurate only within some 10\% in terms of magnetic energy.

In force-free extrapolations the free magnetic energy is only of the order of 10\% of the magnetic energy (see \sect{S:free.energy}).
Considering that the force-free extrapolation can be accurate only to some 10\%, too, this implies that the derived free magnetic energy is not well constrained.
Basically this says that the results from a force-free extrapolation are consistent with a potential field (in terms of energy). Thus one should not use the free energy derived from a force-free extrapolation for a quantitative analysis.

\subsection{When can we trust force-free extrapolations?}\label{S:trust}

We now turn to a quantitative measure of when a force-free extrapolation would give results in terms of free magnetic energy that can be trusted.
For this we define a free-$\beta$, $\beta_{\rm{f}}$, that in analogy to plasma-$\beta$ compares the pressure (or internal energy) to the \emph{free} energy of the magnetic field,
\begin{equation}\label{E:def.free.beta}
\beta_{\rm{f}} = \frac{p}{~(B_{\rm{}}^2-B_{\rm{pot}}^2)\,/\,(2\mu_0)~} ~.
\end{equation}
With the relative free energy $\dE_{\rm{free}}{=}\,(B^2{-}B_{\rm{pot}}^2)\,/B^2$ as defined in \eqn{E:delta.free}, free-$\beta$ reads
\begin{equation}\label{E:beta.free}
\beta_{\rm{f}} ~~=~~ \frac{p}{~\dE_{\rm{free}} ~ B^2/(2\mu_0)~}
        ~~=~~ \frac{\beta}{~\dE_{\rm{free}}~} ~.
\end{equation}
Finally with \eqn{E:err.ff} we find the ratio of the error of the force-free extrapolation (by neglecting the impact of the plasma)
to the relative free energy to be
\begin{equation}\label{E:err.free.energy}
\frac{~\dE_{\rm{err}}~}
     {~\dE_{\rm{free}}~}
 ~~=~~    \frac{3}{2} ~ \beta_{\rm{f}} ~.
\end{equation}
The conclusion is that as long as the free-$\beta$ as defined in \eqn{E:def.free.beta} is of unity or even larger, $\beta_{\rm{f}} \gtrsim 1$, then the error of the force-free extrapolation (because of the presence of the plasma) is at least as big as the difference of the extrapolated field from a potential field. In other words: within the errors the extrapolated field is the same as a potential field (in terms of energy).

If $\beta_{\rm{f}} \gtrsim 1$, then also the Lorentz force is not large compared to the other terms in the momentum equation, but will be balanced by the pressure gradient. The condition for this was in \eqn{E:Efree.beta} that $\dE_{\rm{free}}{\approx}\beta$. Then it follows from \eqn{E:beta.free} that $\beta_{\rm{f}}{\approx}1$ proving the above statement.

Only if $\beta_{\rm{f}}$ is \emph{significantly} below unity, we can trust the magnetic energy (and thus the free energy) derived from a force-free extrapolation. Similarly, only if $\beta_{\rm{f}}$ is significantly below unity we can trust the currents derived from a force-free extrapolation and thus any conclusions drawn on the distribution of (Ohmic) heating in the corona based on this (cf. \app{S:para.currents}).

In general, plasma-$\beta$ ranges from several percent to 10\% (see \sect{S:beta}).
The relative free energy $\dE_{\rm{free}}$ is the order of 10\% for normal active regions and might reach about 50\% in flaring active regions (see \sect{S:free.energy}).
With this, based on \eqn{E:beta.free} the free-$\beta$ is \emph{not} significantly below unity, but typically will be of the order of unity in normal active regions and maybe $\beta_{\rm{f}}{\approx}0.2$ in flaring active regions.
Therefore the trust in the free magnetic energy derived from force-free extrapolations should be limited.

\subsection{Comparison to 3D MHD models}\label{S:beta.mhd}

With our estimations in \sects{S:beta} and \ref{S:free.energy} that plasma-$\beta$ and the relative free energy $\dE_{\rm{free}}$ are of the same order, we would expect the free-$\beta$ is of order unity. While those are only estimations, one can use a 3D MHD accounting for the magnetic field and the plasma for a closer check.

Again, we use the results from the 3D simulation of \cite{Bingert+Peter:2011} to calculate free-$\beta$. For this we perform a potential field extrapolation using the vertical magnetic field from the 3D model at a height of 5\,Mm which is (roughly) the base of the corona in that model.
Together with the actual magnetic field in the MHD model we then calculate the free energy $\dE_{\rm{free}}$ according to \eqn{E:delta.free} in the volume of each grid cell of the computation.
Finally we use the plasma-$\beta$ from the model and evaluate the free-$\beta$ at each grid cell according to \eqn{E:beta.free}.

\begin{figure}
\includegraphics[width=88mm]{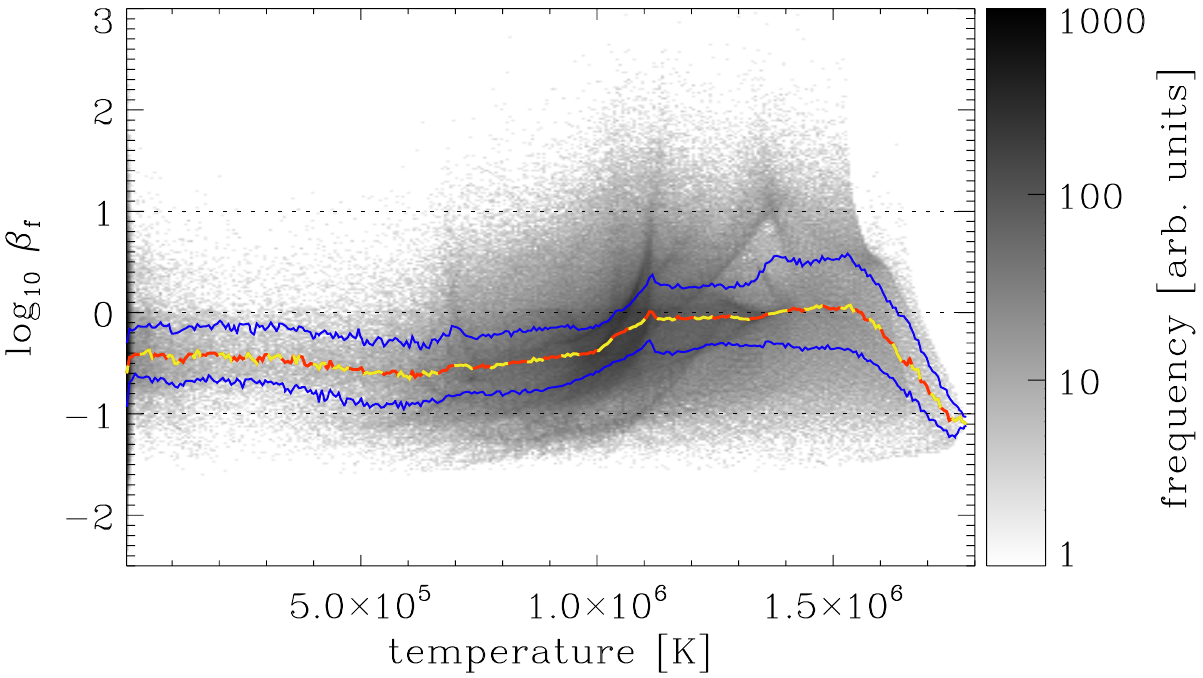}
\caption{Free-$\beta$ in a 3D MHD model. The plot shows a 2D histogram of $\beta_{\rm{f}}$ as defined in \eqn{E:def.free.beta} as a function of the temperature in the computational domain (above 5\,Mm).
The yellow-red dashed line indicates the median variation (here similar to
the mean variation). 
The solid blue lines show the 25 and 75 percentiles, i.e.\ half of the data points
are in-between these lines.
The color bar shows the frequency in the histogram.
Data based on the model presented in \cite{Bingert+Peter:2011}.
See \sect{S:beta.mhd}.
\label{F:free.beta}}
\end{figure}

In \fig{F:free.beta} we show the resulting distribution of $\beta_{\rm{f}}$ in the 3D MHD model.
This demonstrates that in the coronal part of the computational domain, where the temperatures reach values of the order of 1\,MK to 1.5\,MK, free-$\beta$ has values of the order of $\beta_{\rm{f}}\approx1$.
Thus the numerical model confirms the estimation for $\beta_{\rm{f}}$ at the end of \sect{S:trust} for a non-flaring active region.
Consequently, according to \eqn{E:err.free.energy} the the free magnetic energy is of the order of the error in magnetic energy if neglecting the effcets of the plasma. 
This supports our conclusion from \sect{S:trust} that one cannot fully trust force-free extrapolations in terms of free magnetic energy.

\subsection{Accuracy of force-free extrapolations}\label{S:accuracy}

The above discussion does not consider the accuracy of the method to perform the force-free extrapolation. 
Few analytical solutions of highly symmetric and comparably simple magnetic configurations exist \citep[e.g.,][]{Low+Lou:1990,Titov+Demoulin:1999}.
If compared to these, force-free extrapolations can recover the true magnetic energy from the analytical solution only to some 10\% if only the lower boundary, i.e., the surface magnetic field is specified \citep[based on six algorithms investigated by][]{Schrijver+al:2006}.
In particular, different methods ``will not necessarily converge to the correct, or even the same, solution'' \citep{Metcalf+al:2008}.
A comparison of a dozen methods of extrapolation using the same data set of an observed active region magnetic field provides free magnetic energies in the range of 3\% to 25\% \citep{DeRosa+al:2009}. 
This does not reflect a real error because the true magnetic field in the corona is not know in these cases. 
Still, it shows that the scatter between the different methods is considerable and that the error solely introduced by the extrapolation might be at least of the order of 20\%.

It would interesting to directly compare force-free extrapolations to 3D MHD models. For this one could use the magnetic field at the base of the corona in the MHD model to perform the extrapolation. In particular when using different MHD models, e.g.\ for an evolved \citep{Bingert+Peter:2011} or an emerging active region \citep{Chen+al:2014}, the comparison to the magnetic field and the free energy from the extrapolation could provide a further test to the accuracy and reliability of force-free extrapolations. Such a study however, is far beyond the scope of the present investigation.

The arguments in this paper address only the principle difficulty of force-free extrapolations due to the fact that plasma-$\beta$ is not vanishing in the corona.
The errors introduced by the extrapolation algorithms will make things worse.

\section{Possible progress}\label{S:progress}

To overcome the limitations of force-free extrapolations, some progress might be achieved by combining the measured surface magnetic field and the observed loop orientation in the minimisation process \citep{Aschwanden+al:2012,Malanushenko+al:2012,Malanushenko+al:2014}. This would result in solutions where the magnetic energy is not minimised (for the given vector magnetic field at the boundary), but where the magnetic field might be closer to the true field on the Sun.

A direct step forward to account for the effects of the gas pressure would be to perform magneto-hydrostatic (MHS) models \citep[e.g.][]{Ruan+al:2008}.
However such models have the serious shortcoming that they cannot account for the history and evolution of the corona.
This memory of the magnetic field is accounted for by magneto-frictional models \cite[e.g.][]{Cheung+DeRosa:2012}, but they neglect the presence of plasma and essentially face problems similar to the force-free extrapolations discussed here. 
 
To account for the presence of plasma and the history of the magnetic evolution one has to employ 3D MHD models.
Driven by observational data, i.e., the magnetic field at the solar surface as a function of time, such models can represent observed coronal structures, as demonstrated by \cite{Bourdin+al:2013}.
While that model used only line-of-sight magnetograms, the models by \cite{Chen+al:2014,Chen+al:2015} have been driven by the full vector of the magnetic field at the surface, as well as by the velocity field.
Thus full 3D MHD models driven by the observed vector magnetic field are within reach.
At the moment such data-driven 3D MHD models, including global coronal models \citep[e.g.][]{Riley+al:2011}, do not produce fully realistically looking coronae.
One reason for this is that time-step limitations restrict the solar time covered by the models \citep{Peter:2015}.
With further improvements of methods and computational resources, such models might provide the progress needed for a more reliable determination of currents and free energy in the solar corona.

\section{Conclusions\label{S:conclusions}}

We investigated the limitations of force-free extrapolations as a consequence of neglecting the impact of the plasma by assuming plasma-$\beta$ to be negligibly small. We find that observations, models, and theory do not really support this assumption, and that $\beta$ is of the order of a few percent to 10\% or even more.

In our theoretical investigation we define a free-$\beta$, $\beta_{\rm{f}}$, in \eqn{E:def.free.beta}, which is the ratio of the plasma-$\beta$ to  the (relative) free magnetic energy $\dE_{\rm{free}}$. If  $\beta_{\rm{f}} \gtrsim 1$, i.e. if the free energy is \emph{not} substantially larger than plasma-$\beta$, we draw the following conclusions:
\begin{itemize}

\item The error of the force-free extrapolation in terms of magnetic energy is of the same order as the difference of the true magnetic field to a potential field. This is to say that a potential extrapolation is as accurate as a force-free extrapolation.

\item The free magnetic energy derived from the extrapolated field will be within the errors of the force-free extrapolation. Therefore the free magnetic energy derived from a force-free extrapolations cannot be trusted.

\item The currents are not constrained at all in a force-free extrapolation, because they do not have to be parallel to the magnetic field.
 
\end{itemize}
On the Sun and for extrapolations of (non-flaring) active regions in general we provided evidence that free-$\beta$ is of order unity, $\beta_{\rm{f}} \approx 1$. Even in a flaring active region $\beta_{\rm{f}}$ will not be significantly below unity. Thus the above conclusions apply to force-free extrapolations of solar active regions.

Still, one might extract important information on the magnetic connectivity and topology from a force-free extrapolation. However, the quantitative results on the amount of the free magnetic energy available and on the currents to locate the regions of energy deposition cannot really be trusted.
Approaches going beyond current force-free extrapolations are needed (cf.\ \sect{S:progress}) to possibly overcome these limitations.

\begin{acknowledgements}
We would like to thank Sven Bingert and Feng Chen for help in providing data for the comparison to 3D MHD models. 
Sincere thanks are due to Bernd Inhester and Thomas Wiegelmann for discussions on the extrapolations.
We kindly acknowledge comments by Karel Schrijver, in particular his suggestion to give an outlook on possible progress.
We also thank the referee for constructive comments helping to improve the paper.
J.W. acknowledges funding from the People Programme (Marie Curie Actions) of the European Union's Seventh Framework Programme (FP7/2007-2013) under REA grant agreement No. 623609. 
L.P.C. acknowledges funding by the Max-Planck Princeton Center for Plasma Physics.
RHC's contribution was carried out in the context of Deutsche Forschungsgemeinschaft SFB 963 ``Astrophysical Flow Instabilities and Turbulence'' (Project A16).
\end{acknowledgements}



\appendix

\section{Reliability of field-parallel currents in a force-free extrapolation\label{S:para.currents}}

The following discussion will show that if plasma-$\beta$ \eqn{E:beta} is of the same order as the free energy of the magnetic field 
$\dE_{\rm{free}}$ \eqn{E:delta.free}, then the currents \emph{parallel} to the magnetic field in general need \emph{not}  be dominated 
by those of a force-free extrapolation alone.
Following \eqn{E:beta.free} the above condition $\beta\approx\dE_{\rm{free}}$ is equivalent to the free-$\beta$ being of order unity, $\beta_{\rm{f}}{\approx}1$.

In the discussion in \sect{S:force} we show that the Lorentz force due to (small) currents that are perpendicular to the magnetic 
field $\bm{B}$ can be balanced by the plasma as long as $\beta$ is of the order of the free energy \eqn{E:Efree.beta}. %
One can decompose the currents $\bm{j}$ associated with the true magnetic field into a part $\bm{j}_{\rm{ff}}$ associated with the 
magnetic field from the force-free extrapolation that is parallel to $\bm{B}$ and another part $\bm{j}_{\rm{L}}$ that is associated 
with the Lorentz force being balanced by the plasma and that is not necessarily aligned with $\bm{B}$. 
These two parts, $\bm{j}_{\rm{ff}}$ and $\bm{j}_{\rm{L}}$ would correspond to the two options in \eqn{E:ff.options}.
The part balanced by the plasma can be further separated into a component perpendicular $\bm{j}_{\rm{L}}^{\,\perp}$ and a component 
parallel $\bm{j}_{\rm{L}}^{\,||}$ to $\bm{B}$. Thus one can write the currents associated with the true magnetic field as
\begin{equation}\label{E:curr.comp}
\bm{j} = \bm{j}_{\rm{ff}}  +  \bm{j}_{\rm{L}}^{\,\perp}  +  \bm{j}_{\rm{L}}^{\,||} ~.
\end{equation}

The currents of the true magnetic field, $\bm{j}{=}\nabla{\times}\bm{B}/\mu_0$, will be divergence-free, $\nabla\cdot\bm{j}=0$ 
(because the divergence of a curl is zero). Likewise, the currents of the force-free field alone will satisfy $\nabla\cdot\bm{j}_{\rm{ff}}=0$, 
and consequently also $\nabla\cdot(\,\bm{j}_{\rm{L}}^{\,\perp}{+}\bm{j}_{\rm{L}}^{\,||}\,)=0$.
The component $\bm{j}_{\rm{L}}^{\,\perp}$ is determined through the pressure gradient $\nabla p$ balancing the Lorentz force $\bm{j}_{\rm{L}}^{\,\perp}\times\bm{B}$, and in general $\nabla\cdot\bm{j}_{\rm{L}}^{\,\perp}\ne0$.

Assuming that the only parallel currents are those of the force-free extrapolation (i.e. $\bm{j}_{\rm{L}}^{\,||}=0$) would imply  $\nabla\cdot\bm{j}_{\rm{L}}^{\,\perp}=0$, which cannot be met in general.
Therefore the assumption 
$\bm{j}_{\rm{L}}^{\,||}=0$ will not apply --- in general $\bm{j}_{\rm{L}}^{\,||} \ne 0$. 
This argument also suggests that the parallel $\bm{j}_{\rm{L}}^{\,||}$ and perpendicular $\bm{j}_{\rm{L}}^{\,\perp}$ currents associated with the Lorentz force balanced by the plasma are likely to have similar magnitudes (because $\nabla\cdot\bm{j}_{\rm{L}}^{\,||} + \nabla\cdot\bm{j}_{\rm{L}}^{\,\perp} = 0$). 

The current introduced by the non-zero plasma-$\beta$ therefore has components both perpendicular and parallel to the magnetic field,
neither of which  are captured by a force-free extrapolation. In particular this rises doubts for quantitative studies where one uses the 
currents derived from an extrapolation to investigate the location and magnitude of the heating in the corona.



\begin{thebibliography}{46}
\expandafter\ifx\csname natexlab\endcsname\relax\def\natexlab#1{#1}\fi

\bibitem[{{Amari} {et~al.}(2000){Amari}, {Luciani}, {Mikic}, \&
  {Linker}}]{Amari+al:2000}
{Amari}, T., {Luciani}, J.~F., {Mikic}, Z., \& {Linker}, J. 2000, \apjl, 529,
  L49

\bibitem[{{Aschwanden} {et~al.}(2012){Aschwanden}, {Wuelser}, {Nitta}, {Lemen},
  {DeRosa}, \& {Malanushenko}}]{Aschwanden+al:2012}
{Aschwanden}, M.~J., {Wuelser}, J.-P., {Nitta}, N.~V., {et~al.} 2012, \apj,
  756, 124

\bibitem[{{Bingert} \& {Peter}(2011)}]{Bingert+Peter:2011}
{Bingert}, S. \& {Peter}, H. 2011, \aap, 530, A112

\bibitem[{{Bourdin} {et~al.}(2013){Bourdin}, {Bingert}, \&
  {Peter}}]{Bourdin+al:2013}
{Bourdin}, P.-A., {Bingert}, S., \& {Peter}, H. 2013, \aap, 555, A123

\bibitem[{{Chen} \& {Peter}(2015)}]{Chen+Peter:2015}
{Chen}, F. \& {Peter}, H. 2015, \aap, 581, A137

\bibitem[{{Chen} {et~al.}(2014){Chen}, {Peter}, {Bingert}, \&
  {Cheung}}]{Chen+al:2014}
{Chen}, F., {Peter}, H., {Bingert}, S., \& {Cheung}, M.~C.~M. 2014, \aap, 564,
  A12

\bibitem[{{Chen} {et~al.}(2015){Chen}, {Peter}, {Bingert}, \&
  {Cheung}}]{Chen+al:2015}
{Chen}, F., {Peter}, H., {Bingert}, S., \& {Cheung}, M.~C.~M. 2015, Nature
  Physics, 11, 492

\bibitem[{{Cheung} \& {DeRosa}(2012)}]{Cheung+DeRosa:2012}
{Cheung}, M.~C.~M. \& {DeRosa}, M.~L. 2012, \apj, 757, 147

\bibitem[{Cirtain {et~al.}(2013)Cirtain, Golub, Winebarger,
  {et~al.}}]{Cirtain+al:2013}
Cirtain, J.~W., Golub, L., Winebarger, A.~L., {et~al.} 2013, \nat, 493, 501

\bibitem[{{De Moortel} \& {Pascoe}(2009)}]{DeMoortel+Pascoe:2009}
{De Moortel}, I. \& {Pascoe}, D.~J. 2009, \apjl, 699, L72

\bibitem[{{De Rosa} {et~al.}(2009){De Rosa}, {Schrijver}, {Barnes},
  {et~al.}}]{DeRosa+al:2009}
{De Rosa}, M.~L., {Schrijver}, C.~J., {Barnes}, G., {et~al.} 2009, \apj, 696,
  1780

\bibitem[{{Edwin} \& {Roberts}(1983)}]{Edwin+Roberts:1983}
{Edwin}, P.~M. \& {Roberts}, B. 1983, \solphys, 88, 179

\bibitem[{{Gary}(2001)}]{Gary:2001}
{Gary}, G.~A. 2001, \solphys, 203, 71

\bibitem[{{Guo} {et~al.}(2015){Guo}, {Erd{\'e}lyi}, {Srivastava}, {Hao},
  {Cheng}, {Chen}, {Ding}, \& {Dwivedi}}]{Guo+al:2015}
{Guo}, Y., {Erd{\'e}lyi}, R., {Srivastava}, A.~K., {et~al.} 2015, \apj, 799,
  151

\bibitem[{Lagg {et~al.}(2004)Lagg, Woch, Krupp, \& Solanki}]{Lagg+al:2004}
Lagg, A., Woch, J., Krupp, N., \& Solanki, S.~K. 2004, \aap, 414, 1109

\bibitem[{Lin {et~al.}(2004)Lin, Kuhn, \& Coulter}]{Lin+al:2004}
Lin, H., Kuhn, J.~R., \& Coulter, R. 2004, \apj, 613, L177

\bibitem[{Lin {et~al.}(2000)Lin, Penn, \& Tomczyk}]{Lin+al:2000}
Lin, H., Penn, M.~J., \& Tomczyk, S. 2000, \apj, 541, L83

\bibitem[{{Low} \& {Lou}(1990)}]{Low+Lou:1990}
{Low}, B.~C. \& {Lou}, Y.~Q. 1990, \apj, 352, 343

\bibitem[{{Malanushenko} {et~al.}(2014){Malanushenko}, {Schrijver}, {DeRosa},
  \& {Wheatland}}]{Malanushenko+al:2014}
{Malanushenko}, A., {Schrijver}, C.~J., {DeRosa}, M.~L., \& {Wheatland}, M.~S.
  2014, \apj, 783, 102

\bibitem[{{Malanushenko} {et~al.}(2012){Malanushenko}, {Schrijver}, {DeRosa},
  {Wheatland}, \& {Gilchrist}}]{Malanushenko+al:2012}
{Malanushenko}, A., {Schrijver}, C.~J., {DeRosa}, M.~L., {Wheatland}, M.~S., \&
  {Gilchrist}, S.~A. 2012, \apj, 756, 153

\bibitem[{{Metcalf} {et~al.}(2008){Metcalf}, {De Rosa}, {Schrijver}, {Barnes},
  {van Ballegooijen}, {Wiegelmann}, {Wheatland}, {Valori}, \&
  {McTtiernan}}]{Metcalf+al:2008}
{Metcalf}, T.~R., {De Rosa}, M.~L., {Schrijver}, C.~J., {et~al.} 2008,
  \solphys, 247, 269

\bibitem[{{Nakariakov} \& {Ofman}(2001)}]{Nakariakov+Ofman:2001}
{Nakariakov}, V.~M. \& {Ofman}, L. 2001, \aap, 372, L53

\bibitem[{{Nakariakov} \& {Verwichte}(2005)}]{Nakariakov+Verwichte:2005}
{Nakariakov}, V.~M. \& {Verwichte}, E. 2005, Living Reviews in Solar Physics,
  2, 3

\bibitem[{{Peter}(2015)}]{Peter:2015}
{Peter}, H. 2015, Phil. Trans. R. Soc. A, 373, 20150055

\bibitem[{{Peter} {et~al.}(2012{\natexlab{a}}){Peter}, {Abbo}, {Andretta},
  {Auch{\`e}re}, {Bemporad}, {Berrilli}, {Bommier}, {Braukhane}, {Casini},
  {Curdt}, {Davila}, {Dittus}, {Fineschi}, {Fludra}, {Gandorfer}, {Griffin},
  {Inhester}, {Lagg}, {Degl'Innocenti}, {Maiwald}, {Sainz}, {Pillet},
  {Matthews}, {Moses}, {Parenti}, {Pietarila}, {Quantius}, {Raouafi},
  {Raymond}, {Rochus}, {Romberg}, {Schlotterer}, {Sch{\"u}hle}, {Solanki},
  {Spadaro}, {Teriaca}, {Tomczyk}, {Bueno}, \& {Vial}}]{Peter+al:2012}
{Peter}, H., {Abbo}, L., {Andretta}, V., {et~al.} 2012{\natexlab{a}},
  Experimental Astronomy, 33, 271

\bibitem[{{Peter} \& {Bingert}(2012)}]{Peter+Bingert:2012}
{Peter}, H. \& {Bingert}, S. 2012, \aap, 548, A1

\bibitem[{{Peter} {et~al.}(2012{\natexlab{b}}){Peter}, {Bingert}, \&
  {Kamio}}]{Peter+al:2012.loop}
{Peter}, H., {Bingert}, S., \& {Kamio}, S. 2012{\natexlab{b}}, \aap, 537, A152

\bibitem[{{Peter} {et~al.}(2006){Peter}, {Gudiksen}, \&
  {Nordlund}}]{Peter+al:2006}
{Peter}, H., {Gudiksen}, B.~V., \& {Nordlund}, {\AA}. 2006, \apj, 638, 1086

\bibitem[{Priest(1982)}]{Priest:1982}
Priest, E.~R. 1982, Solar Magnetohydrodynamics (D. Reidel, Dordrecht)

\bibitem[{{Reale}(2014)}]{Reale:2014}
{Reale}, F. 2014, Living Reviews in Solar Physics, 11, 4

\bibitem[{{Riley} {et~al.}(2011){Riley}, {Lionello}, {Linker}, {Mikic},
  {Luhmann}, \& {Wijaya}}]{Riley+al:2011}
{Riley}, P., {Lionello}, R., {Linker}, J.~A., {et~al.} 2011, \solphys, 274, 361

\bibitem[{{Rosner} {et~al.}(1978){Rosner}, {Tucker}, \&
  {Vaiana}}]{Rosner+al:1978}
{Rosner}, R., {Tucker}, W.~H., \& {Vaiana}, G.~S. 1978, \apj, 220, 643

\bibitem[{{Ruan} {et~al.}(2008){Ruan}, {Wiegelmann}, {Inhester}, {Neukirch},
  {Solanki}, \& {Feng}}]{Ruan+al:2008}
{Ruan}, P., {Wiegelmann}, T., {Inhester}, B., {et~al.} 2008, \aap, 481, 827

\bibitem[{Schrijver {et~al.}(2006)Schrijver, Derosa, Metcalf,
  {et~al.}}]{Schrijver+al:2006}
Schrijver, C.~J., Derosa, M.~L., Metcalf, T.~R., {et~al.} 2006, \solphys, 235,
  161

\bibitem[{{Schrijver} \& {van
  Ballegooijen}(2005)}]{Schrijver+vanBallegooijen:2005}
{Schrijver}, C.~J. \& {van Ballegooijen}, A.~A. 2005, \apj, 630, 552

\bibitem[{Solanki {et~al.}(2003)Solanki, Lagg, Woch,
  {et~al.}}]{Solanki+al:2003}
Solanki, S.~K., Lagg, A., Woch, J., {et~al.} 2003, \nat, 425, 692

\bibitem[{{Sun} {et~al.}(2012){Sun}, {Hoeksema}, {Liu}, {Wiegelmann},
  {Hayashi}, {Chen}, \& {Thalmann}}]{Sun+al:2012}
{Sun}, X., {Hoeksema}, J.~T., {Liu}, Y., {et~al.} 2012, \apj, 748, 77

\bibitem[{{Thalmann} {et~al.}(2012){Thalmann}, {Pietarila}, {Sun}, \&
  {Wiegelmann}}]{Thalmann+al:2012}
{Thalmann}, J.~K., {Pietarila}, A., {Sun}, X., \& {Wiegelmann}, T. 2012, \aj,
  144, 33

\bibitem[{{Thalmann} {et~al.}(2014){Thalmann}, {Tiwari}, \&
  {Wiegelmann}}]{Thalmann+al:2014}
{Thalmann}, J.~K., {Tiwari}, S.~K., \& {Wiegelmann}, T. 2014, \apj, 780, 102

\bibitem[{{Thalmann} \& {Wiegelmann}(2008)}]{Thalmann+Wiegelmann:2008}
{Thalmann}, J.~K. \& {Wiegelmann}, T. 2008, \aap, 484, 495

\bibitem[{{Titov} \& {D{\'e}moulin}(1999)}]{Titov+Demoulin:1999}
{Titov}, V.~S. \& {D{\'e}moulin}, P. 1999, \aap, 351, 707

\bibitem[{{van Ballegooijen} {et~al.}(2011){van Ballegooijen}, {Asgari-Targhi},
  {Cranmer}, \& {DeLuca}}]{vanBallegooijen+al:2011}
{van Ballegooijen}, A.~A., {Asgari-Targhi}, M., {Cranmer}, S.~R., \& {DeLuca},
  E.~E. 2011, \apj, 736, 3

\bibitem[{{White} \& {Verwichte}(2012)}]{White+Verwichte:2012}
{White}, R.~S. \& {Verwichte}, E. 2012, \aap, 537, A49

\bibitem[{White(2005)}]{White:2005}
White, S.~M. 2005, in Chromospheric and Coronal Magnetic Fields, Vol. ESA
  SP--596, 10

\bibitem[{{Wiegelmann} \& {Sakurai}(2012)}]{Wiegelmann+Sakurai:2012}
{Wiegelmann}, T. \& {Sakurai}, T. 2012, Living Reviews in Solar Physics, 9, 5

\bibitem[{{Wiegelmann} {et~al.}(2014){Wiegelmann}, {Thalmann}, \&
  {Solanki}}]{Wiegelmann+al:2014}
{Wiegelmann}, T., {Thalmann}, J.~K., \& {Solanki}, S.~K. 2014, \aapr, 22, 78

\end{thebibliography}
\end{document}